\documentclass[aps,
prl,
twocolumn,
floatfix,
superscriptaddress,10pt,longbibliography
]{revtex4-2}
\usepackage{amssymb,amsmath,amsthm,graphicx,color}
\usepackage{bm}
\usepackage{mathrsfs}
\usepackage{enumitem}
\usepackage{booktabs}
\usepackage{tabularx}
\usepackage{cancel}
\usepackage{pifont}
\usepackage{soul}
\usepackage[colorlinks]{hyperref}
\usepackage{mathtools,mathrsfs,esint}
\usepackage{array}
\usepackage{xcolor}

\hypersetup{
     breaklinks=true,
    pdfstartview={FitH},    
    colorlinks=true,       
    linkcolor=blue,          
    citecolor=red,        
    filecolor=magenta,      
    urlcolor=blue,           
    anchorcolor=green,      
    linktocpage=true
}

\newcolumntype{C}[1]{>{\centering\arraybackslash}p{#1}}

\newcommand{\mpl}{m_\text{Pl}}

\allowdisplaybreaks

\newcommand{\PRLsection}[1]{\emph{#1.---}}

\begin{document}

\title{{\bf Black hole binary mergers in dynamical Chern-Simons gravity with regularisation}}

\author{Llibert Arest\'e Sal\'o}
\email[]{llibert.areste@uib.cat}
\affiliation{Departament de F\'isica, Universitat de les Illes Balears, Palma de Mallorca, E-07122, Spain}
\affiliation{Institute of Applied Computing \& Community Code (IAC3), Universitat de les Illes Balears, Palma de Mallorca, E-07122, Spain}

\author{Maxence Corman}
\email[]{m.corman@qmul.ac.uk}
\author{Pau Figueras}
\email[]{p.figueras@qmul.ac.uk}
\author{\'Aron D. Kov\'acs}
\email[]{a.kovacs@qmul.ac.uk}
\author{Shunhui Yao}
\email[]{s.yao@qmul.ac.uk}

\affiliation{Centre for Geometry, Analysis and Gravitation, School of Mathematical Sciences, Queen Mary University of London, Mile End Road, London E1 4NS, United Kingdom}


\begin{abstract}
We study the non-linear dynamics of Chern-Simons gravity with numerical relativity for the first time. We use the regularisation prescription to obtain a well-posed initial-value formulation in the gauges commonly used in numerical relativity, and we simulate head-on collisions and binary black hole mergers in the regime of validity of effective field theory, extracting the corresponding scalar and gravitational waves. This opens up the possibility to test general effective field theories of gravity, including those with higher-than-second-order equations of motion, with gravitational waves. 
\end{abstract}

\maketitle

\PRLsection{{\bf Introduction}} The LIGO-Virgo-KAGRA collaboration has made nearly 400 gravitational wave detections since 2015~\cite{LIGOScientific:2026wfs}. The increased sensitivity of the detectors, together with the detection of events with a very high signal-to-noise ratio such as GW250114, has allowed to carry out more precise tests of general relativity (GR) with gravitational waves (GWs)~\cite{LIGOScientific:2025wao,LIGOScientific:2025rid}. This situation will only get better in the near future; with the planned upgrades of the current detectors, and the forthcoming next generation of ground-based~\cite{ET:2019dnz,Evans:2021gyd} and space-based~\cite{LISA_paper} detectors, it will be possible to carry out even more precise tests of GR with GWs. However, the vast majority of tests that are routinely performed assume GR. Although GR has passed all observational tests so far, learning about the fundamental nature of gravity requires comparing observations with the predictions of well-motivated alternatives. Such comparisons could either link possible observational anomalies to specific new physics or place constraints on deviations from GR. 

Effective field theory (EFT) provides a systematic framework to parameterise deviations from GR~\cite{Endlich:2017tqa,Burgess:2020tbq}. EFTs of gravity can reliably describe the strong non-linear dynamics of sufficiently large black holes (BHs), as long as spacetime curvatures stay well below the ultraviolet cut-off scale. Fundamental UV-consistency requirements (such as unitarity, locality, and causality) impose bounds on the Wilson coefficients of the EFTs (see e.g. \cite{Adams:2006sv,deRham:2022hpx}). Exploring the non-linear dynamics of theories both within and outside the viable parameter spaces determined by these positivity bounds is essential for identifying smoking gun signatures that can link GW observations directly to new UV physics. Until recently, however, studies on the full non-linear dynamics of EFTs were mostly restricted to special theories with second-order equations where suitable gauge conditions 
\cite{Kovacs:2019jqj,Kovacs:2020pns,Kovacs:2020ywu,AresteSalo:2022hua,AresteSalo:2023mmd,mccz4} enabled non-linear binary black hole (BBH) merger simulations~\cite{Corman:2022xqg,AresteSalo:2022hua,AresteSalo:2023mmd,Doneva:2023oww,Corman:2024cdr,Corman:2025wun} and effective-one-body models~\cite{Julie:2024fwy}.

Among gravitational EFTs, dynamical Chern-Simons (dCS) gravity is of particular interest~\cite{Alexander:2009tp}. While this theory may be motivated from high energy physics considerations, it arises as the leading higher-derivative parity-violating EFT correction to Einstein's gravity coupled to a (pseudo-)scalar field \cite{Weinberg:2008hq}. Therefore, dCS theory introduces new strong field phenomenology absent in GR, thus offering a unique opportunity to constrain it both with GW observations \cite{Nair:2019iur,Perkins:2021mhb} and EFT consistency principles \cite{Xu:2024iao,Cassem:2026aib,Serra:2022pzl}. Rotating stationary BHs have been constructed in this theory~\cite{Yunes:2009hc,Yagi:2012ya,Stein:2014xba,McNees:2015srl,Lam:2025fzi}, together with the corresponding quasinormal modes~\cite{Wagle:2021tam,Chung:2025gyg}, as well as post-Newtonian calculations of compact binary inspirals~\cite{Sopuerta:2009iy,Yagi:2011xp,Yagi:2012vf,Yagi:2013mbt}. However, so far, the dynamics of this theory has only been explored perturbatively and/or in the probe approximation~\cite{Okounkova:2017dCS,Okounkova:2017yby,Okounkova:2018pql,Okounkova:2019dfo,Okounkova:2019zjf,Richards:2023xsr}. The reason is that the equations of motion of dCS are of third order in derivatives of the spacetime metric, and it does not seem to be possible to find a well-posed formulation (which would allow a fully non-linear evolution) with traditional methods~\cite{Delsate:2014hba}.

In fact, generic EFTs of gravity have higher-than-second-order equations of motion, and therefore the mathematical difficulties encountered in finding a well-posed formulation for dCS gravity are not unique to this specific theory. Recently,~\cite{Figueras:2024bba} introduced a new approach to covariantly regularise a general class of gravitational EFTs, including those with higher-than-second-order equations of motion, leading to a well-posed initial-value formulation of such theories (see also \cite{Cayuso:2017iqc,Cayuso:2017iqc,Cayuso:2023xbc,Lara:2021piy,Lara:2026dwh,Lara:2026xcs} for alternative methods). This opens up the possibility of exploring the non-linear regime of a large class of theories with numerical simulations. In this paper, we generalise the covariant regularisation method to dCS theory and use it to simulate head-on collisions of spinning BHs and BBH mergers, extracting the corresponding GWs, which can then be compared with the predictions from GR. 

\PRLsection{Conventions} We use Latin letters $a,b,c,\ldots$ to denote abstract indices, follow the curvature conventions of~\cite{Wald:1984rg}, and set $c=1$.

\PRLsection{{\bf Methods}} The bare action for the most general scalar-tensor theory of gravity up to four derivatives of the fields is~\cite{Weinberg:2008hq}
\begin{equation}
    \begin{aligned}
    S=&~\int d^4x\sqrt{-g}\Big\{
        \tfrac{\mpl^2}{2}\, R -\tfrac{1}{2}\,(\nabla_a\phi)(\nabla^a\phi) - V(\phi) \\
     &\hspace{0.5cm}+\ell^2\Big(\mpl\,F(\phi)\,\mathscr{L}_\text{GB}+\mpl\,G(\phi)\,\tilde R_{abcd}\,R^{abcd}\\
     &\hspace{0.5cm}+\tfrac{1}{4\,\mpl^2}\, H(\phi)\big( (\nabla_a\phi)(\nabla^a\phi) \big)^2\Big)
    \Big\}\,,
    \end{aligned}
    \label{eq:action0}
\end{equation}
where $V(\phi)$ is the scalar potential,
\begin{equation}
    \mathscr{L}_\text{GB} = R^2 - 4\, R_{ab}\,R^{ab}+R_{abcd}\,R^{abcd}\,
\end{equation}
is the Gauss-Bonnet term, $\tilde R_{abcd}=\tfrac{1}{2}\,\epsilon_{ab}^{\phantom{ab}ef}R_{efcd}$
is the dual Riemann tensor, $\mpl^2=\frac{1}{8\pi\,G}$, $\ell$ is the length scale associated with the UV cut-off and $F(\phi)$, $G(\phi)$ and $H(\phi)$ are smooth functions of their arguments. The action \eqref{eq:action0} includes Einstein-scalar-Gauss-Bonnet (EsGB) gravity ($G(\phi)=H(\phi)=0$), dynamical Chern-Simons (dCS) gravity ($F(\phi)=H(\phi)=0$) and a particular k-essence theory ($F(\phi)=G(\phi)=0$). All of these theories have received considerable attention in the literature, but so far only for the case $G(\phi)=0$ it has been possible to study their non-linear dynamics. In this paper, we uncover the fully non-linear regime of dCS gravity for the first time.

The equations of motion that follow from varying the action \eqref{eq:action0} are
\begin{align}
    \Box\phi =&~V'(\phi) \nonumber\\
    &+\ell^2\bigg\{\frac{H(\phi)}{\mpl^2}\left[ (\nabla\phi)^2\Box\phi + 2\,(\nabla^a\phi)(\nabla^b\phi)\nabla_a\nabla_b\phi\right]\nonumber\\
    &+\tfrac{3}{4\,\mpl^2}\,H'(\phi)\big((\nabla^a\phi)(\nabla_a\phi)\big)^2\nonumber\\
    &-\mpl\left( F'(\phi)\,\mathscr{L}_\text{GB}+G'(\phi)\, \tilde R_{abcd}\,R^{abcd}\right)\bigg\}\,, \label{eq:original_phi_evo}\\
    \mpl^2\,G_{ab} =&~ (\nabla_a\phi)(\nabla_b\phi)+g_{ab}\bigl(V(\phi)-\tfrac{1}{2}\,(\nabla\phi)^2\bigr) \nonumber\\
    &- 8\,\mpl\,\ell^2\nabla^c\nabla^d\left[F(\phi)\,\tilde{\tilde{R}}_{d(ab)c} -G(\phi)\,\tilde R_{d(ab)c}\right]\,, \label{eq:original_g_evo}
\end{align}
where $G_{ab}$ is the Einstein tensor associated to the spacetime metric $g_{ab}$ and $\tilde{\tilde{R}}_{abcd}=\frac{1}{2}\epsilon_{cd}{}^{ef}\tilde{R}_{abef}$ is the double-dual of the Riemann tensor. The equations of motion \eqref{eq:original_phi_evo}--\eqref{eq:original_g_evo} contain up to three derivatives of the spacetime metric $g_{ab}$ and up to two derivatives of the scalar field $\phi$.
To obtain a well-posed initial-value formulation of \eqref{eq:action0} and simulate BBH mergers, we implement the covariant regularisation method of~\cite{Figueras:2024bba} to the theory \eqref{eq:action0}.

Following~\cite{Figueras:2024bba}, we consider perturbative field redefinitions proportional to the 0$^\text{th}$-order equations of motion, 
\begin{align}
    g_{ab}\to&~ g_{ab}+\alpha\,\ell^2\Big[R_{ab}-\tfrac{1}{\mpl^2}\bigl((\nabla_a\phi)(\nabla_b\phi) +g_{ab}\,V(\phi)\bigr)\Big]\,, \label{eq:redef_metric}\\
    \phi\to&~ \phi + \gamma\,\ell^2(\Box\phi-V'(\phi))\,. \label{eq:redef_scalar}
\end{align}
This gives rise to a new action that agrees with \eqref{eq:action0} modulo terms of order $O(\ell^4)$. The new, ``regularised'' action is given by
\begin{equation}
    \begin{aligned}
        S_\text{reg}&~=\int d^4x\sqrt{-g}\Big\{\tfrac{\mpl^2}{2}\, R -\tfrac{1}{2}\,(\nabla_a\phi)(\nabla^a\phi) -V(\phi) \\
     &+\ell^2\Big(\mpl\,F(\phi)\,\mathscr{L}_\text{GB}+\mpl\,G(\phi)\,\tilde R_{abcd}\,R^{abcd}\\
     &\hspace{1.25cm}+\tfrac{1}{4\,\mpl^2}\, H(\phi)\big( (\nabla_a\phi)(\nabla^a\phi) \big)^2\Big) \\
     &+\alpha\,\ell^2\Big[
     \Big(\mpl^2\,R^{ab}-2\bigl((\nabla^a\phi)(\nabla^b\phi)+V(\phi)\,g^{ab}\bigr)\Big)G_{ab} \\
     &\hspace{0.8cm}+\tfrac{1}{2\,\mpl^2}\,\bigl(\big((\nabla_a\phi)(\nabla^a\phi)\big)^2-4\,V(\phi)(\nabla_a\phi)(\nabla^a\phi)\\
     &\hspace{2.5cm}-8\,V(\phi)^2\bigr)
     \Big]\\
     &+ \gamma\,\ell^2\,\bigl(\phi\,\Box^2\phi-V'(\phi)(2\,\Box\phi-V'(\phi))\bigr)
     \Big\}\,,
    \end{aligned}
    \label{eq:Sreg}
\end{equation}
where we have redefined $\alpha\to -2\,\alpha$. The parameters $\alpha$ and $\gamma$ in the field redefinitions \eqref{eq:redef_metric}--\eqref{eq:redef_scalar} are dimensionless numbers that, at this stage, can be chosen freely. The actions \eqref{eq:Sreg} and \eqref{eq:action0} describe the same physics in a regime where EFT is valid (see e.g.,~\cite{Burgess:2020tbq}).

The equations of motion that follow from varying \eqref{eq:Sreg} can be reduced to a strongly hyperbolic system of coupled (second-order) non-linear wave equations. Standard hyperbolic PDE theory implies that the initial value problem for such a system of PDEs is at least locally well-posed for sufficiently regular initial data~\footnote{Local well-posedness holds e.g. with respect to a Sobolev norm adapted to the auxiliary variables, see also~\cite{Thaalba:2026eyv}.}. To obtain such a reduced system, we introduce the new variables
\begin{align}
    \phi^{(1,0)}_a\equiv&~\nabla_a\phi\,,\\
    \phi^{(0,1)}\equiv&~\Box\phi-V'(\phi)\,, 
    \label{eq:residual_scalar_def}\\
    \bar R_{ab}\equiv&~ R_{ab}[g]-\mathcal{S}_{ab}\,, 
    \label{eq:ricci_residual_def}
\end{align}
where $R_{ab}[g]$ is the Ricci tensor expressed in terms of the metric and its derivatives up to second order (after suitable gauge-fixing), and
\begin{equation}
    \mathcal{S}_{ab}\equiv\tfrac{1}{\mpl^2}\left(\phi^{(1,0)}_a\phi^{(1,0)}_b+g_{ab}\,V(\phi)\right)\,.
    \label{eq:def_S}
\end{equation}
Then, the equations of motion for the theory~\eqref{eq:Sreg} can be reduced to the system consisting of \eqref{eq:residual_scalar_def}, \eqref{eq:ricci_residual_def}, and the evolution equations for the auxiliary variables $\phi^{(1,0)}_a$, $\phi^{(0,1)}$ and $\bar R_{ab}$ given in the Appendix, eqs.~\eqref{eq:evo_phi_a}--\eqref{eq:evo_Ric_bar}. The fields $\phi^{(0,1)}$ and $\bar R_{ab}$ may be interpreted as UV regulators since they satisfy massive wave equations with masses around (or beyond) the UV cut-off~\footnote{In fact, $\bar R_{ab}$ contains a scalar (the trace) and a tensor (the traceless part) degrees of freedom.}, with asymptotic masses (squared) given by $m_S^2=-\frac{1}{2\gamma\ell^2}$ and $m_T^2=-\frac{1}{2\alpha\ell^2}$, respectively~\footnote{If the scalar potential $V(\phi)$ has a mass term, it will also contribute to the asymptotic value of the scalar mass squared $m_S^2$ and the bound on $\gamma$ will change accordingly.}. Therefore, to avoid a tachyonic instability in the weakly coupled regime and achieve stable evolutions, we need to impose $\alpha,\,\gamma<0$. In practice and for the results presented below, we choose $\alpha=\gamma=-\frac{1}{2}$, unless otherwise stated. We have also considered other values of these parameters, and the dependence of the observables on $\alpha$ and $\gamma$ for different choices of $\ell$ is addressed in the Appendix.

We solve the evolution equations  of the regularised theory numerically with two independent codes, one that uses generalised harmonic coordinates (GHC)~\cite{Pretorius:2004jg,Pretorius:2006tp}, expanding on the numerical scheme of~\cite{East:2011aa,East:2023nsk}, and another one that uses the CCZ4 formulation of the Einstein equations for the metric and the standard moving-punctures gauge~\cite{Alic:2011gg} based on \texttt{GRChombo}~\cite{Andrade:2021rbd}. More details on the numerical schemes, convergence results, and quantitative comparisons between the two codes and with the results of the decoupling limit can be found in the Appendix.

\PRLsection{{\bf Results}}
In this paper we focus on the non-linear dynamics of BHs in dCS gravity $(F(\phi)=H(\phi)=0$ in \eqref{eq:action0}) and present some additional results in EsGB theory (in the modified generalised harmonic gauge and the regularised formulation) for comparison. We restrict to a shift-symmetric (pseudo-)scalar with $V(\phi)=0$ and $G(\phi)=\lambda\,\phi$ (analogously $F(\phi)=\lambda\,\phi$ for EsGB), where $\lambda$ is the dimensionless coupling; this choice guarantees that rotating BHs in this theory develop scalar hair. We measure the length scales in terms of the total ADM mass $M$ and fix $\mpl=1/\sqrt{8\pi G}=1$.

\begin{figure}[t]
\includegraphics[width=\columnwidth]{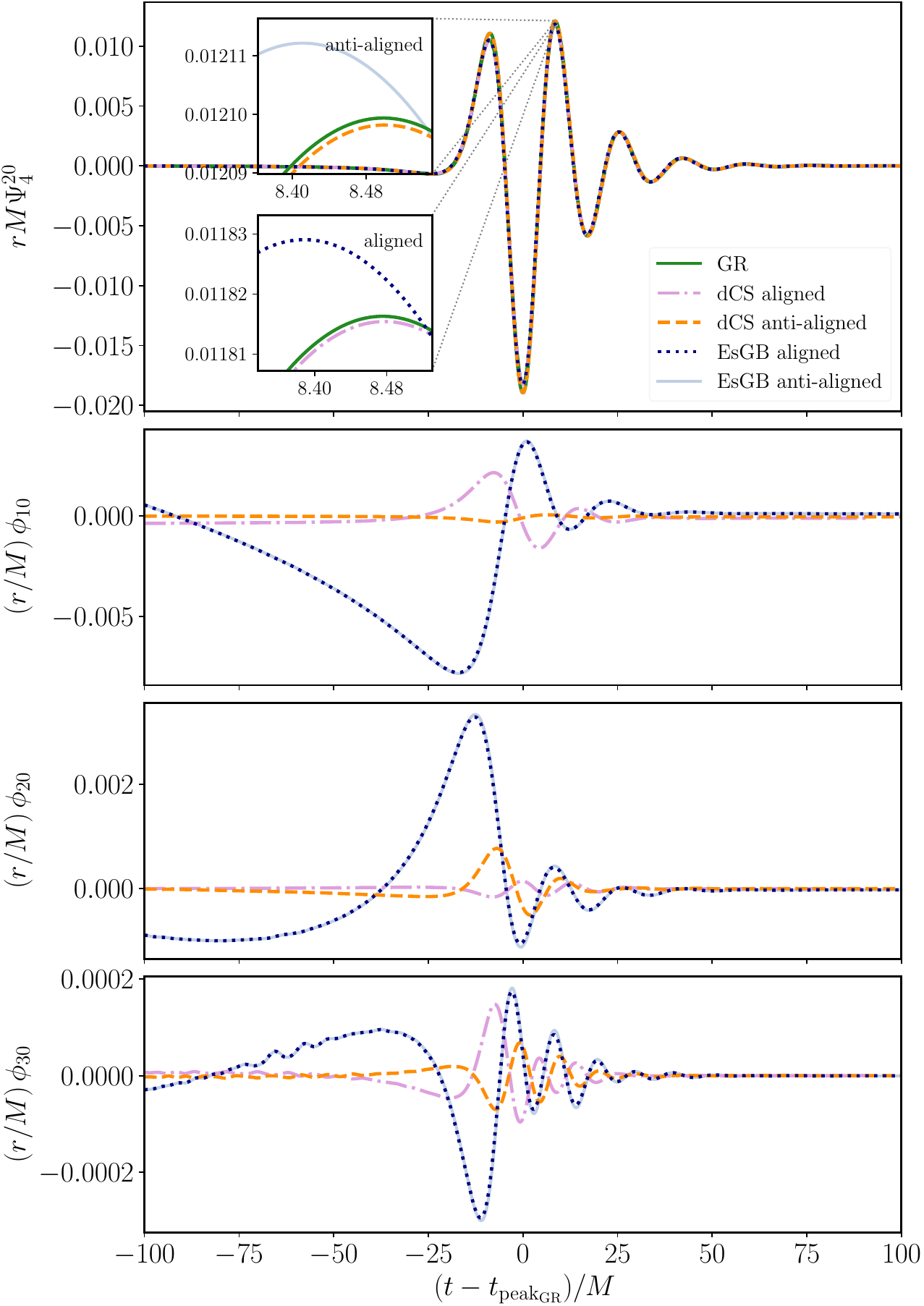}
\caption{Head-on collisions of BHs in dCS, EsGB gravity and GR with mass ratio 1.5 and dimensionless spins of 0.62 in aligned and anti-aligned configurations: From top to bottom, we show $r\,M \Psi_4^{20}$, $(r/M)\phi_{10}$, $(r/M)\phi_{20}$ and $(r/M)\phi_{30}$ extracted at $R=100M$. The time is shifted such that the GR waveform peaks at $t=0$. There is a clear distinction between dCS and EsGB in the emitted scalar radiation for aligned and anti-aligned spin configurations.
}\label{fig:headon_aligned}
\end{figure}

\paragraph{Head-on collisions.} First, we consider the head-on collisions of spinning BHs. The results presented here use GHCs (see Fig.~\ref{fig:headon_aligned} and Appendix for more details on the numerical set-up). We set up initial data that solves the constraints of vacuum GR using the conformal thin sandwich method described in Ref.~\cite{East:2012zn}. The scalar field and its time derivative, as well as the auxiliary fields $\phi^{(1,0)}_a$, $\phi^{(0,1)}$ and $\bar R_{ab}$ vanish initially. We choose an initial separation of $80M$, and set their initial velocities to the values corresponding to the two BHs being marginally bound at infinite separation, a mass ratio of 1.5, and (anti-)aligned intrinsic dimensionless spins of $0.62$. We fix the UV length scale such that $\ell^2= 0.01 M^2$, $\alpha=-3.5$, $\gamma=-1.0$, and fix the coupling to $\lambda\,\ell^2 = 0.04 m_2^2$, where $m_2$ is the mass of the smallest BH. The dimensionless spin of the remnant BH of (anti-)aligned head-on collision is $(0.12)\,0.32$. With this set-up, BHs develop scalar hair over a period of $30M$ and the merger occurs at around $t=350M$, so there is a sufficiently long time separation between the growth of the scalar hair and the merger.  

The results are displayed in Fig.~\ref{fig:headon_aligned}, showing the $(2,0)$ mode of the Weyl scalar $\Psi_4$ and the scalar $(1,0)$, $(2,0)$ and $(3,0)$ modes, extracted at a finite radius of $R= 100M$ and shifted in time such that the GR waveform peaks at $t=0$. We also show the corresponding modes assuming GR and shift-symmetric EsGB with the same intrinsic parameters. We find that for the collisions in dCS theory, $\Psi_4$ exhibits a slightly lower peak amplitude than in GR and EsGB, with no appreciable advance or delay of the merger. This is because these head-on collisions happen too quickly for the effects of the dCS coupling to accumulate significantly. Conversely, EsGB mergers occur noticeably earlier for both spin configurations (see \cite{Corman:2025wun} for a detailed comparison of BBH mergers in GR and in EsGB). Regarding scalar radiation, dCS and EsGB are clearly distinguishable at equal coupling: EsGB produces larger scalar amplitudes and a clear dephasing relative to dCS. Finally, in EsGB theory the scalar radiation is nearly identical for aligned and anti-aligned spins, in contrast to dCS where there is a clear distinction between different spin orientations.

\paragraph{Black hole binary inspirals.}
We next consider quasi-circular inspirals of equal-mass BHs using the CCZ4 code, see the link for an animation~\footnote{\texttt{https://www.youtube.com/watch?v=Th0Pi4ue4Vk}}; analogous results obtained with the GHC code can be found in the Appendix. For the results shown here, both BHs have intrinsic dimensionless spins of $0.4$, with one spin aligned and the other anti-aligned with respect to the orbital angular momentum. We set the individual puncture ADM masses to $m_1=m_2=0.5M$ and the initial coordinate separation to $10M$. We fix the dCS coupling to $\lambda\ell^2=0.01m_2^2$ and choose $\ell^2=0.01M^2$. We construct GR puncture initial data with \texttt{TwoPunctures}~\cite{Ansorg:2004ds,Paschalidis:2013oya} and compute the initial momenta from post-Newtonian expressions using the \texttt{NRPyPN} library~\cite{NRPyPN}. The scalar field and its conjugate momentum initially vanish, as do the auxiliary fields $\phi^{(1,0)}_a$, $\phi^{(0,1)}$ and $\bar R_{ab}$. For these simulations, the size of the domain is $L=512M$ and the resolution in the coarsest level is $dx=2M$; we add 7 additional levels of refinement with a $2:1$ ratio. 

\begin{figure}[t]
\includegraphics[width=\columnwidth]{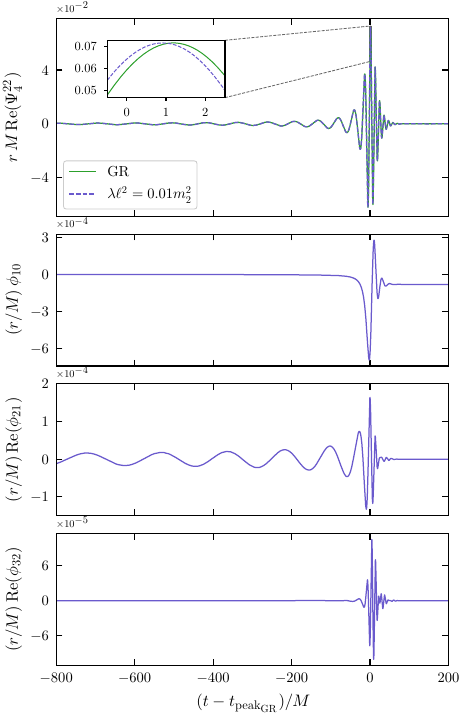}
\caption{Gravitational and scalar waveforms from an equal-mass binary with dimensionless spins of magnitude 0.4 anti-aligned with each other, $\lambda\,\ell^2=0.01 m_2^2$ with $\ell^2=0.01M^2$, as well as GR for comparison: From top to bottom, we show $r\,M\,\text{Re}(\Psi_4^{22})$, $(r/M)\phi_{10}$, $(r/M)\text{Re}(\phi_{21})$ and $(r/M)\text{Re}(\phi_{32})$ extracted at $R=100M$. The time coordinate is shifted so that the GR waveform peaks at $t=0$ (see a \href{https://youtu.be/Th0Pi4ue4Vk}{3D~movie of the simulation}).
}
\label{fig:inspiral_waveforms}
\end{figure}

Figure~\ref{fig:inspiral_waveforms} shows the $(2,2)$ mode of $\Psi_4$ in dCS and GR, as well as the scalar $(1,0)$, $(2,1)$ and $(3,2)$ modes extracted at $R=100M$. The gravitational waveforms in dCS and GR are close, as expected,
since the dCS correction to the metric enters at $O(\ell^4/M^4)$,
 and the discrepancy becomes apparent only in the
zoomed inset near the peak.
 The scalar $(2,1)$ mode oscillates during inspiral and decays after merger, whereas the $(1,0)$ mode exhibits a pronounced merger transient and settles to a non-zero late-time value at this finite extraction radius. This value arises here because the two BHs have equal masses and anti-aligned spins, so their individual scalar charges at infinity cancel during the pre-merger phase, but the remnant BH is spinning and therefore it carries a non-vanishing scalar hair. We can qualitatively compare with~\cite{Okounkova:2017dCS}, bearing in mind that their results were obtained by evolving the leading-order scalar field on a dynamical GR BBH background (and hence without backreaction~\footnote{Note that~\cite{Okounkova:2019zjf} considers the backreaction perturbatively but suffers from secular growth.}). In their case, the two BHs have aligned spins, so a non-vanishing scalar dipole is present both before and after the merger, as can be seen in their Figure 1. The oscillatory behaviour of the $(2,1)$ scalar mode and the burst of the $(1,0)$ mode are qualitatively similar, in spite of the different intrinsic spin configurations. See the Appendix for the waveforms for a BBH merger with aligned spins obtained with the GHC code.

\paragraph{Ghost instability.} The EFT is expected to reliably describe a system as long as its characteristic time/length scale (i.e., the radius of the smallest BH in our setting) is large compared to $\ell$. Nevertheless, the regularised dCS theory~\eqref{eq:Sreg} contains massive regulator fields, some of which are ghosts ($\phi^{(0,1)}$ and the traceless part of $\bar R_{ab}$). On the other hand, the masses of these ghost fields are $O(1/\ell)$ (for suitable choices of $\alpha$ and $\gamma$). For consistent initial data compatible with the regime of validity of EFT, the ghost fields are excited with negligible amplitudes initially. When such data are evolved by the non-linear regularised equations, the massive ghost fields are expected to eventually decay~\cite{Figueras:2025gal,Gavassino:2026xjw}. However, at strong coupling, these interactions can change the character of the regulator fields, turning them into tachyons. This leads to runaway behaviour and the breakdown of the Cauchy evolution, signalling the breakdown of the EFT since the higher-derivative terms neglected in the action~\eqref{eq:action0} become large. In this section, we show how this ghost instability develops in strongly coupled (regularised) dCS theory~\eqref{eq:Sreg}.

We consider the non-linear evolution of a single BH of mass $M$ and dimensionless spin of $0.7$. We fix the asymptotic masses of the ghost fields by choosing $\alpha=\gamma=-\frac{1}{2}$, $\ell^2=0.04\,M^2$, and we vary the dimensionless dCS coupling $\lambda$ to increase/decrease the strength of the interactions. The source functions $H^a$ are set to be constant in time to maintain Kerr-Schild-like coordinates. To monitor the development of the ghost instability, we compute the averages of $\phi^{(0,1)}$ and the spacetime Ricci scalar $R\equiv g^{ab}R_{ab}=g^{ab}(\bar R_{ab}+\mathcal{S}_{ab})$ on the apparent horizon (AH) as
\begin{align}
    \langle \phi^{(0,1)^2} \rangle _{\rm AH} =&~ \frac{1}{\mathcal{A}}\int_{\rm AH} \phi^{(0,1)^2} dA \,, \label{eq:av_phi01}\\
    \langle R \rangle _{\rm AH} =&~ \frac{1}{\mathcal{A}}\int_{\rm AH} |R| \ dA \,, \label{eq:av_Rscalar}
\end{align}
where $\mathcal{A}$ is the area of the AH. 

\begin{figure}[t]
\includegraphics[width=\columnwidth]{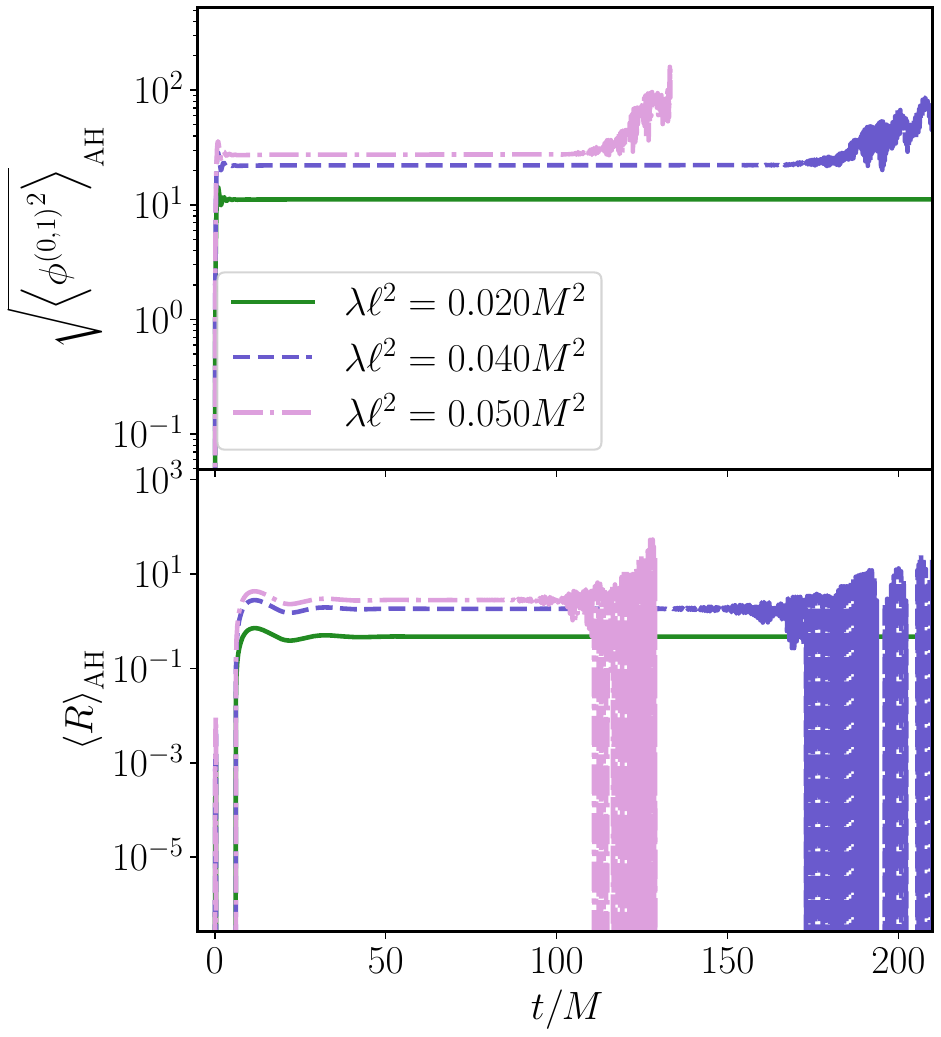}

\caption{Ghost instability for a single BH of ADM mass $M$ and dimensionless spin 0.7. We show $\langle \phi^{(0,1)^2} \rangle _{\rm AH}$ and $\langle R\rangle_{\rm AH}$, see eqs.~\eqref{eq:av_phi01}--\eqref{eq:av_Rscalar}, as functions of time for the couplings $ \lambda\,\ell^2 = \left(0.02,0.04,0.05\right)M^2$ with $\ell^2 = 0.04 M^2$ and $\alpha=\gamma=-\frac{1}{2}$ fixed. For larger dCS coupling, the ghost instability develops faster but it switches off for sufficiently small coupling (at least for the duration of the simulation $t/M=1600$).}\label{fig:ghost}
\end{figure}

The results depicted in Fig.~\ref{fig:ghost} show that for fixed values of the asymptotic masses of the ghost fields, the instability develops faster for larger couplings, whereas no instability is observed (at least for the duration of the simulation) for sufficiently small couplings. This indicates that the instability is only triggered at strong couplings (relative to the size of the BH) that lie outside the regime of validity of the EFT. Interestingly, in Fig.~\ref{fig:ghost} the instability in the scalar regulator ($\phi^{(0,1)}$) appears to set in slightly later than the growth in the Ricci scalar ($R$). One possible explanation is that, as the background evolves, first-derivative couplings act as effective anti-damping terms (as in \cite{Figueras:2025gal}) and begin to amplify curvature perturbations before they amplify scalar perturbations. We leave a more careful analysis of this to future work~\footnote{Note that these results are not in contradiction with the head-on collisions reported earlier, where no instability is observed; they correspond to different choices of initial dimensionless spin, $\ell$ and $\lambda$.}. 

\PRLsection{{\bf Conclusions}} In this paper we studied collisions of spinning BHs in dCS gravity with numerical relativity, solving the theory fully non-linearly for the first time. Although the original theory \eqref{eq:action0} leads to third-order equations of motion, covariant regularisation~\cite{Figueras:2024bba} allows us to obtain an equivalent EFT with a well-posed initial-value problem which can be solved. Furthermore, we have shown that, for initial data within the regime of validity of the EFT, it is possible to solve the full non-linear theory consistently with the massive regulator (ghost) fields decaying. 

In a companion paper~\cite{Long_paper}, we will use the methods developed here to carry out more detailed comparisons between the waveforms produced by GR and dCS theory for various BBH configurations and couplings. The aim is to identify smoking-gun features in the waveforms that can be used to discriminate between GR and dCS gravity. 

This paper is a proof of principle of the power of regularisation to extract the unique non-linear physics of a particular EFT of gravity of physical interest, even if it has higher-than-second-order equations of motion (and hence ghost degrees of freedom). However, most tests of GR with GWs assume vacuum gravity. Therefore, it would be of interest to apply the regularisation approach~\cite{Figueras:2024bba} to an EFT of vacuum gravity. In four spacetime dimensions, the leading higher-derivative corrections to the Einstein-Hilbert action are cubic in the spacetime curvature, and have equations of motion that are fourth order in derivatives of the metric. We will report the results for this theory elsewhere.

\PRLsection{Acknowledgments} We thank Will East, Nils Siemonsen and Katy Clough for useful discussions. We thank Chloe Richards and collaborators for sharing the data used for comparison to decoupling approximation with us.
MC and ADK acknowledge support from the Simons Foundation International and the Simons Foundation through Simons Foundation grant SFI-MPS-BH-00012593-13. We thank the GRTL collaboration (\url{www.grtlcollaboration.org}) for their support and code development work. 
Calculations were performed on the Urania HPC system at the Max Planck Computing and Data Facility. 
This work used the DiRAC Memory Intensive service Cosma8 at Durham University, managed by the Institute for Computational Cosmology on behalf of the STFC DiRAC HPC Facility (\url{www.dirac.ac.uk}). The DiRAC service at Durham was funded by BEIS, UKRI and STFC capital funding, Durham University and STFC operations grants. DiRAC is part of the UKRI Digital Research Infrastructure. 
We acknowledge as well the EuroHPC Joint Undertaking for awarding this project access to the EuroHPC supercomputer LUMI-C, hosted by CSC (Finland) and the LUMI consortium through a EuroHPC
Extreme Scale Access call, under project EHPC-EXT-2025E02-070.

\bibliography{references}

\clearpage

\onecolumngrid
\begin{center}
\textbf{\large Appendix}
\end{center}
\vspace{0.5\baselineskip}
\twocolumngrid

\begin{figure}[t!]
\includegraphics[width=0.98\columnwidth]{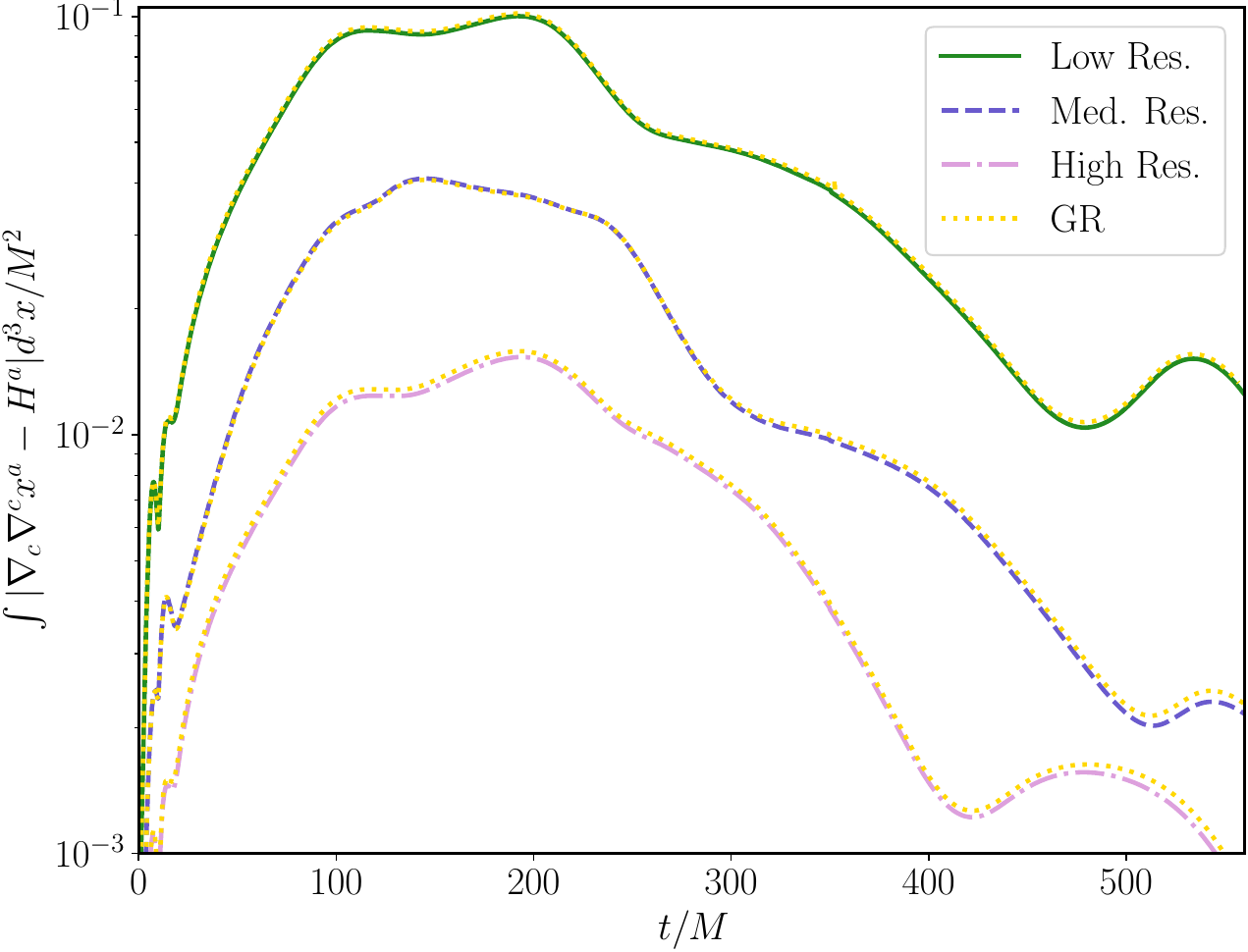}
\includegraphics[width=0.98\columnwidth]{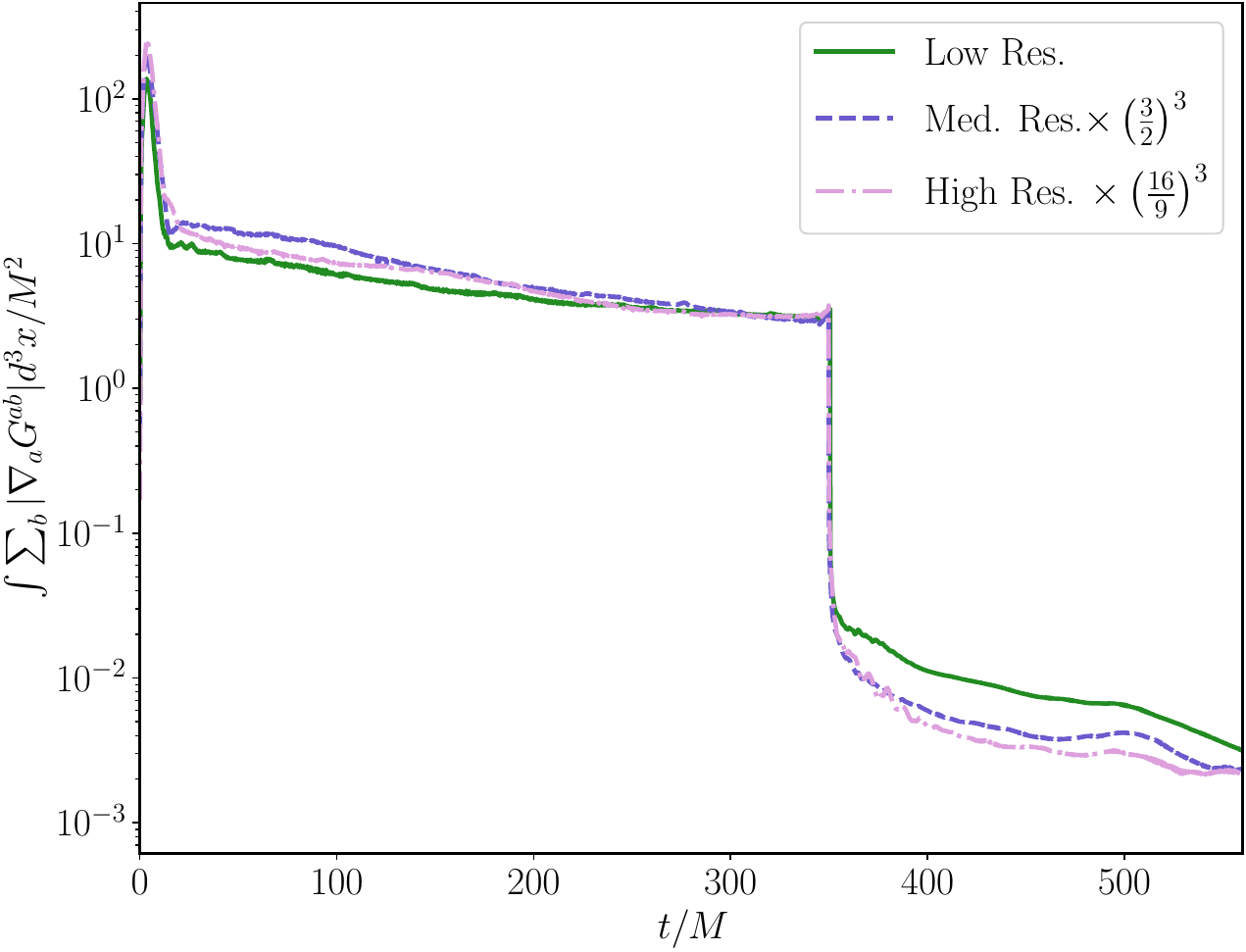}
\vspace{-0.1cm}
\caption{Convergence of the constraints with increasing resolution of a head-on collision of BHs in dCS gravity with mass ratio 1.5, aligned dimensionless spins of 0.62, a UV length scale of $\ell^2=0.01 M^2$ and a dimensionless coupling $\lambda$ such that $ \lambda\,\ell^2 = 0.04 m_2^2$, where $m_2$ is the ADM mass of the smallest BH. We show both the norm of the generalised harmonic constraints $C^a\equiv H^a-\Box x^a=0$ (top) and the norm of the Bianchi constraint $\nabla_a G^{ab}=0$ (bottom)  integrated over the computational domain. For the former we also show the constraints for the same system evolved in GR. The constraints converge at roughly third order.}\label{fig:headon_convergence_cnst}
\vspace{-0.5cm}
\end{figure}

\PRLsection{{\bf Numerical convergence of generalised harmonic scheme}}
Here we provide results for the convergence tests for the numerical implementation of the regularised equations in the GHC code~\cite{Pretorius:2006tp,East:2011aa,East:2023nsk} that we use. 
The lowest resolution has seven levels of mesh refinement with a refinement ratio $2:1$, and a grid spacing of $dx \sim 0.005M$ on the finest level. The medium and high resolutions are $\frac{3}{2}$ and $\frac{16}{9} \times$ as high,
respectively. 
We consider the head-on collision presented in the main text as a representative case.
We construct initial data in harmonic coordinates and evolve it in the damped harmonic gauge~\cite{Lindblom:2009tu,Choptuik:2009ww}, which has been found to work well for a large
number of highly dynamical spacetimes.  In the top panel of Fig.~\ref{fig:headon_convergence_cnst}, we show the integrated generalised harmonic constraint violation $|C_a|$ as a function of coordinate time for
the head-on collisions in dCS gravity  as well as for GR (dotted curve) at several resolutions, demonstrating that it exhibits third order convergence.
Although we use fourth-order finite difference
stencils and RK4 time integration, this level of convergence is consistent with the third-order interpolation in time used to set values on the boundaries of the mesh refinement levels. In the bottom panel, we show the norm of the Bianchi constraint violation
$\sum_b |\nabla_a G^{ab}|$ integrated over domain, for the dCS evolutions. The constraint violation converges to zero with increasing resolution at third order before
merger and fourth order thereafter.  

\begin{figure}[t!]
\includegraphics[width=0.98\columnwidth]{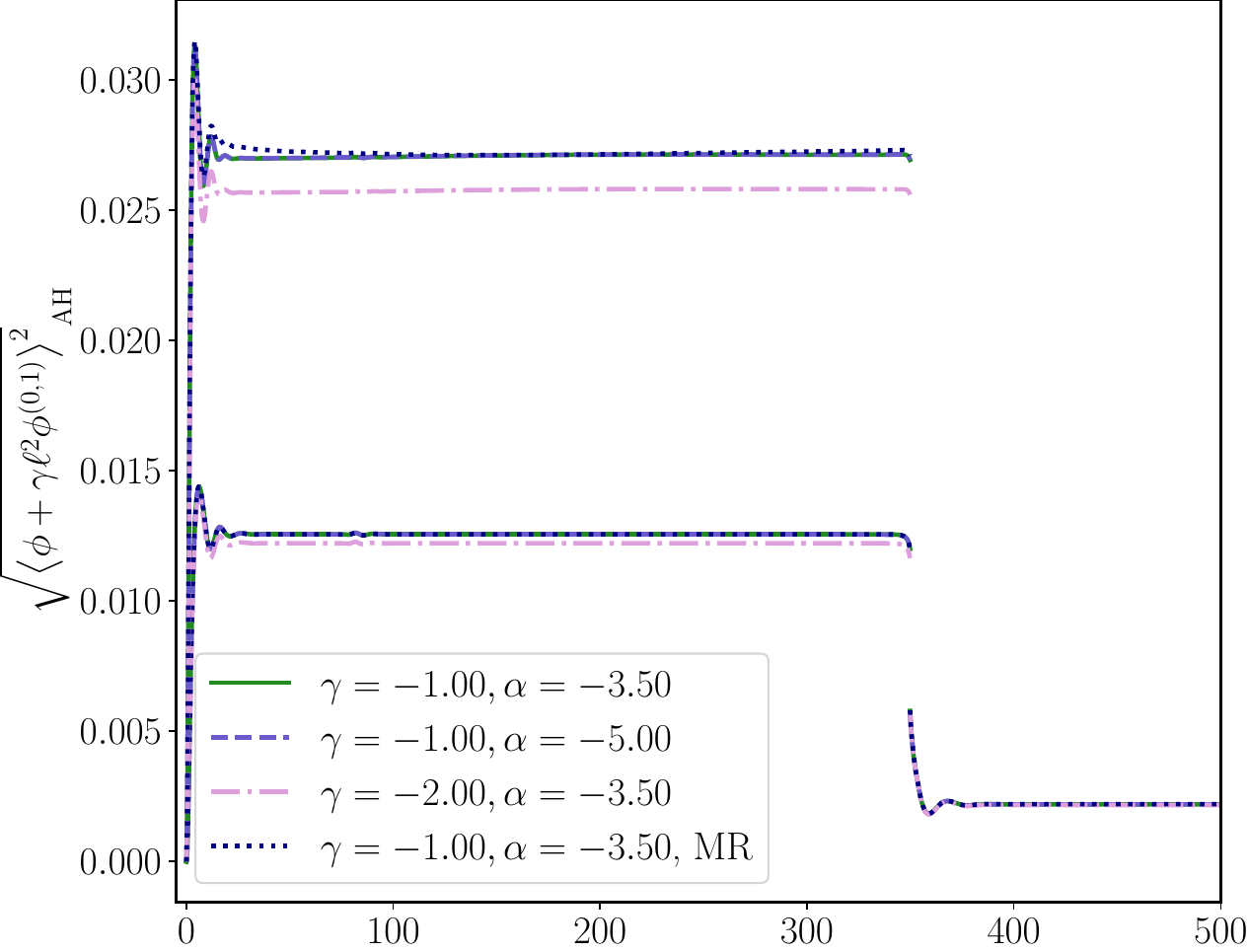}
\includegraphics[width=0.98\columnwidth]{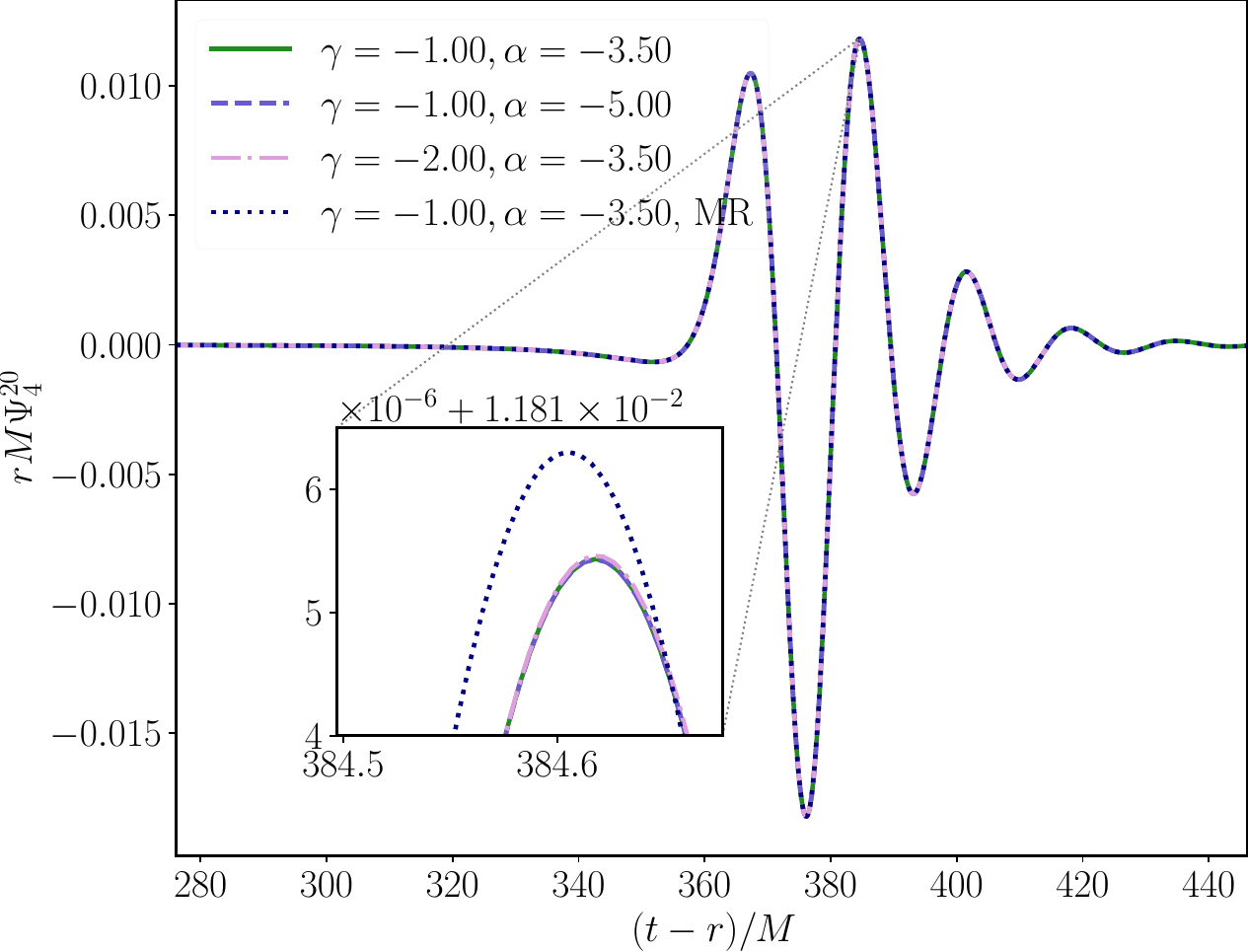}
\vspace{-0.1cm}
\caption{Impact of varying $\alpha$ and $\gamma$ parameters in a head-on collision of BHs in dCS gravity with mass ratio 1.5, dimensionless spin of 0.62 aligned, a UV length scale of $\ell^2=0.01 M^2$ and a dimensionless coupling $\lambda$ such that $ \lambda\,\ell^2 = 0.04 m_2^2$. We show both the original scalar field on horizon (top) and the (2,0) mode of the Weyl scalar extracted at a finite radius as a function of retarded time. For the latter we also show the waveform at a different resolution. The impact of $\alpha$ and $\gamma$ on the waveform is below truncation error.
}\label{fig:headon_alpha_gamma_dependence}
\vspace{-0.5cm}
\end{figure}

\PRLsection{{\bf Dependence on the regularisation frame}} In Fig.~\ref{fig:headon_alpha_gamma_dependence} we test the dependence of the results on the head-on collisions presented in the main text 
on the choice of parameters $\alpha$ and $\gamma$ that appear in the field redefinitions \eqref{eq:redef_metric}--\eqref{eq:redef_scalar}. Considering the same resolutions as in the convergence studies, the top panel shows the original 
scalar field, l.h.s. of \eqref{eq:redef_scalar}, averaged over the AH for different values of $\alpha$ and $\gamma$ at the highest resolution (bottom curves) as well as at medium resolution (top curves).  Similarly, the bottom panel shows the $(2,0)$ mode of $\Psi_4$ extracted at $R=100M$ as a function of retarded time for the same range of values of $\alpha$ and $\gamma$ at high resolution, and the same values of these parameters as in the main text for a medium resolution run. In the regime of validity of the EFT, the results for different $\alpha$ and $\gamma$ are expected to differ
at  $O(\ell^4)$, which is the error of the truncated EFT. In particular, in this regime, these differences should be below the truncation error, as can be observed in the bottom panel of Fig.~\ref{fig:headon_alpha_gamma_dependence}. We note that in the regime of validity of the EFT, the difference between the original scalar field and the evolved scalar field is also controlled by the error of the EFT and, hence, it is below the truncation error too. We also observe a qualitatively similar dependence of the results on $\alpha$ and $\gamma$ for the CCZ4 code.

\begin{figure}[t!]
\includegraphics[width=0.98\columnwidth]{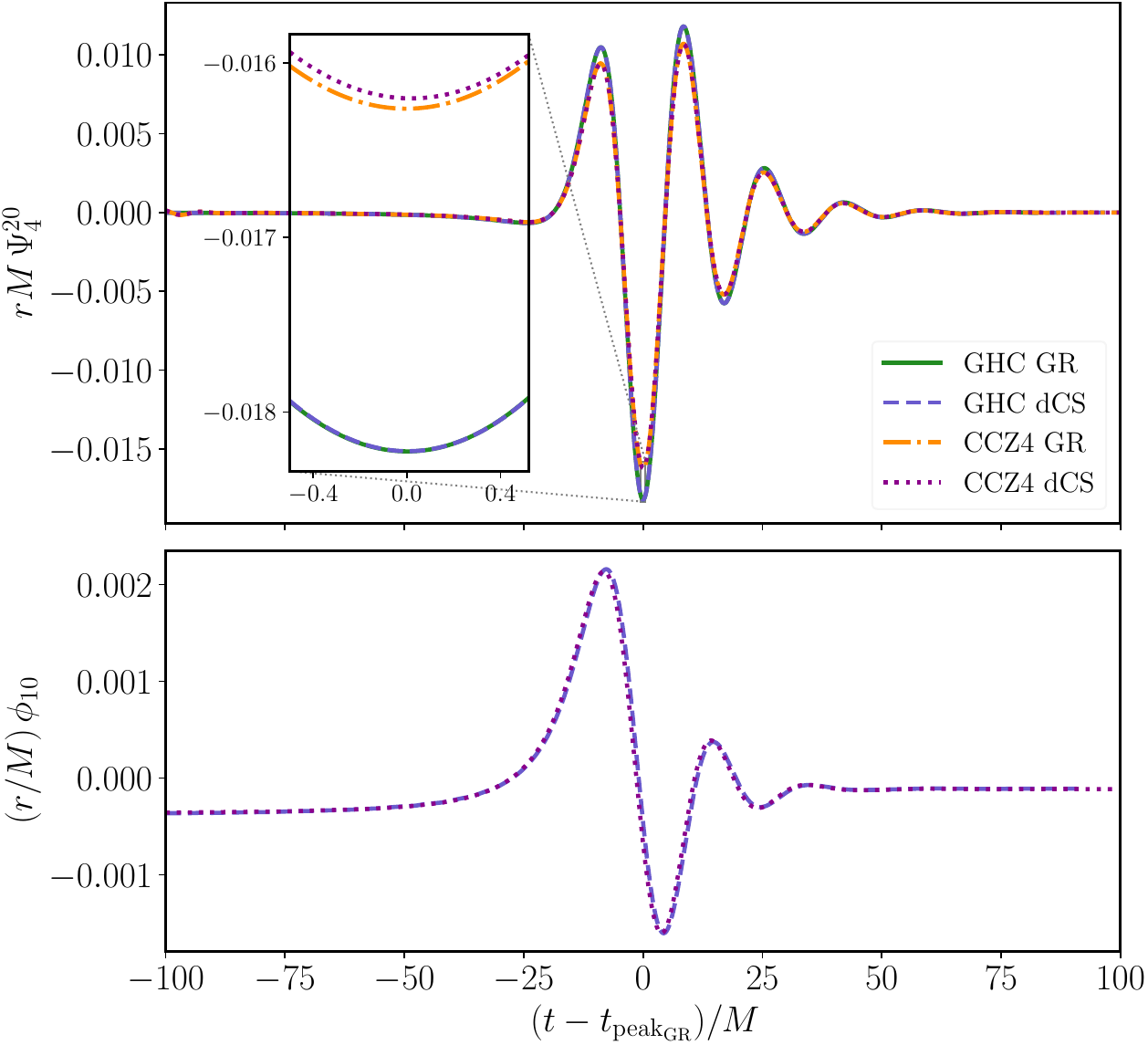}
\vspace{-0.1cm}
\caption{Code comparison of head-on collisions in GR and dCS gravity. We display the $(2,0)$ mode of $\Psi_4$ and the $(1,0)$ mode of the evolved scalar field $\phi$ obtained with both the GHC and CCZ4 codes using the same set-up as in the aligned spins of Fig.~\ref{fig:headon_aligned}, finding a good agreement.
}\label{fig:comparison_head_on_codes}
\end{figure}

\PRLsection{{\bf Code comparison}}
Here we provide a comparison of the two independent numerical implementations used in
this work, the GHC and CCZ4 formulations, for evolving the non-perturbative regularised dCS equations of motion. The codes differ in various ways, including choice of gauge, formulation, initial data construction, boundary conditions and treatment of BHs, making their comparison non-trivial. Details on how the two numerical implementations differ
can be found in Appendix B of~\cite{Corman:2025wun} and references therein. 
We validate the codes using the head-on collision studied in the main text (see Fig.~\ref{fig:headon_aligned}). 
The top panel of Fig.~\ref{fig:comparison_head_on_codes} displays the $(2,0)$ mode of $\Psi_4$ for both GR and dCS with $\lambda\,\ell^2 = 0.04m_2^2$, and the bottom panel shows $(1,0)$ scalar mode, both showing good agreement between the two codes. We have shifted the waveforms in time so that for both codes the merger happens at $t=0$ in GR. We note that the different amplitude at merger in GR likely comes from differences in the initial data (harmonic vs isotropic coordinates) and the slight deviation between dCS and GR is consistent with truncation error.

\begin{figure}[t!]
\includegraphics[width=0.98\columnwidth]{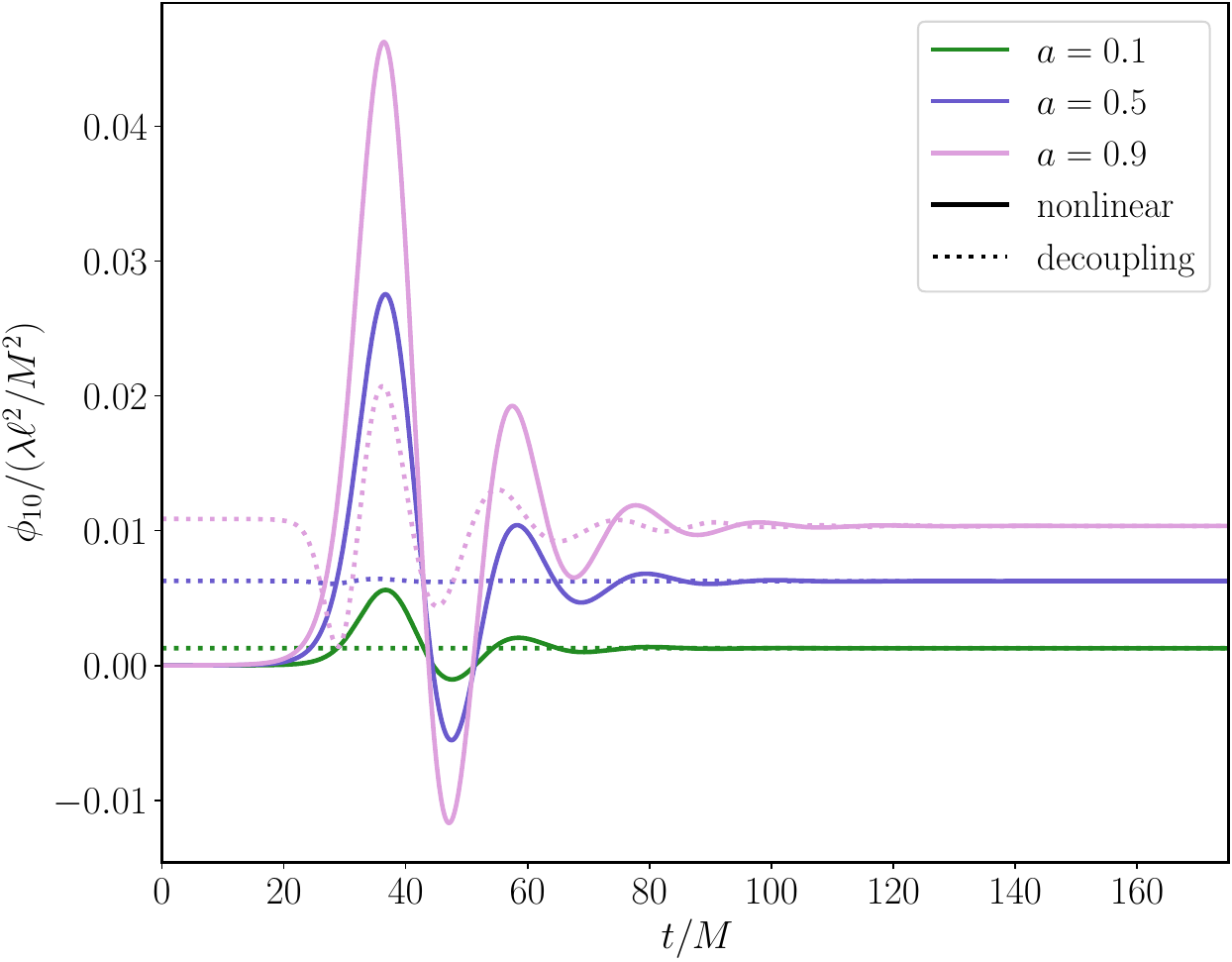}
\vspace{-0.1cm}
\caption{Single BH scalarisation in dCS gravity. Time evolution of $(1,0)$ scalar mode extracted at $R=20M$ rescaled by the coupling parameter, for different values of the (dimensionless) spin. The solid curves were obtained by solving the fully non-linear equations of motion and the dotted lines by solving the equations in the decoupling approximation. The discrepancy at early times is due to different choices of initial data, but the agreement at late times is excellent.
}\label{fig:phi_decoupling_cmp}
\vspace{-0.5cm}
\end{figure}

\PRLsection{{\bf Comparison with the literature}}
In this section, we compare our results with existing results in the literature. The latter have only been obtained in the decoupling approximation, in which the background is determined by GR and the backreaction of
the scalar onto the metric is neglected~\cite{Richards:2023xsr,Richards:2025ows}. 
We focus on the scalarisation of isolated spinning BHs in dCS gravity for different dimensionless spin
values, ranging from $a=0.1$ to $a=0.9$ and a fixed coupling value of $\lambda\,\ell^2=0.01M^2$. In Fig.~\ref{fig:phi_decoupling_cmp} we compare the $(1,0)$ scalar mode extracted at a radius  of $R=20 M$ rescaled by $\lambda\,\ell^2$. We find that the values at late times agree to within $0.2\%$ for all spins. The discrepancy at early times is due to different choices of initial data. Therefore, we conclude that for small enough couplings, the fully backreacted results agree well with the results obtained in the decoupling limit.

\newpage

\PRLsection{{\bf Equations of motion for the regularised theory}} Below we list the evolution equations for the auxiliary variable $\phi^{(1,0)}_a$ and the dynamical variables $\phi^{(0,1)}$ and $\bar R_{ab}$. In their respective evolution equations, \eqref{eq:evo_phi_a}--\eqref{eq:evo_Ric_bar}, the Ricci tensor $R_{ab}$ is expressed in terms of the evolution variables $\bar R_{ab}$ and $\phi^{(1,0)}_a$ via \eqref{eq:ricci_residual_def}, so no new higher-derivative terms are introduced; similarly, the spacetime Weyl tensor $C_{abcd}$ is computed from the spacetime metric $g_{ab}$ and $R_{ab}$.

\begin{widetext}
\begin{align}
    \Box\phi^{(1,0)}_a=&~\nabla_a\phi^{(0,1)}
    +\left(R_a{}^b+V''(\phi)\delta_a{}^b\right)\phi^{(1,0)}_b
    -\mathcal{D}_a\,,\label{eq:evo_phi_a}\\
    \left(\Box+\tfrac{1}{2\,\gamma\,\ell^2}-V''(\phi)\right)\phi^{(0,1)}
    =&~\tfrac{H(\phi)+2\alpha}{2\gamma\mpl^2}
    \left[(\phi^{1,0})^2\phi^{(0,1)}
    +2\,\phi^{(1,0)a}\phi^{(1,0)b}\nabla_a\phi^{(1,0)}_b\right]\nonumber\\
    &+\tfrac{3\,H'(\phi)}{8\gamma\mpl^2}\left((\phi^{(1,0)})^2\right)^2
    +\tfrac{V'(\phi)H(\phi)}{2\gamma\mpl^2}(\phi^{(1,0)})^2
    -\tfrac{\mpl\,}{2\gamma}\left[F'(\phi)\mathscr{L}_{\rm GB}
    +G'(\phi)\tilde C_{abcd}C^{abcd}\right]\nonumber\\
    &-\tfrac{2\alpha}{\gamma}\left[
    R^{ab}\nabla_a\phi^{(1,0)}_b
    +\left(\tfrac{V(\phi)}{\mpl^2}-\tfrac12\,R\right)\phi^{(0,1)}
    -\tfrac{V(\phi)V'(\phi)}{\mpl^2}\right]\,,
    \label{eq:evo_phibox}\\
    \Big(\Box+\tfrac{1}{2\,\alpha\,\ell^2}\Big)\bar R_{ab}=&~2 \,R_{a}{}^{c} R_{bc} -  \tfrac{1}{3}\, R_{ab} R -  \tfrac{1}{2}g_{ab}\big( R_{cd} R^{cd} -\tfrac{1}{6}\,\mathcal{R}^2\big) - 2\, R^{cd} \,C_{acbd} \nonumber\\
        &+\tfrac{1}{\mpl^2}\Big[2\,\nabla_{c}\phi^{(1,0)}_{a}\,\nabla^{c}\phi^{(1,0)}_{b} -2\,\phi^{(0,1)}\nabla_{(a}\phi^{(1,0)}_{b)} +2\,\phi^{(1,0)\,c}\, R_{c(a}\,\phi^{(1,0)}_{b)}  \nonumber\\
        &\hspace{1.2cm}
        - \tfrac{2}{3}\, R\Big(\phi^{(1,0)}_a\,\phi^{(1,0)}_b- \tfrac{1}{4}\, g_{ab} \, (\phi^{(1,0)} )^2\Big)
        + 2\, C_{acbd} \,\phi^{(1,0)\,c}\,\phi^{(1,0)\,d}  
        -2\,R_{ab}V(\phi)\nonumber\\        &\hspace{1.25cm}
        +V'(\phi)\,g_{ab}\bigl(\phi^{(0,1)}+V'(\phi)\bigr)
        +V''(\phi)\bigl(2\,\phi^{(1,0)}_a\,\phi^{(1,0)}_b+g_{ab}\,(\phi^{1,0})^2\bigr)\Big]\nonumber\\
        &+\tfrac{1}{\mpl^4}\Big[(\phi^{(1,0)})^2 \Big(
        - \phi^{(1,0)}_a\,\phi^{(1,0)}_b  + \tfrac{1}{4}\, g_{ab} (\phi^{(1,0)})^2 \Big)+2\,V(\phi)\bigl(\phi^{(1,0)}_a\,\phi^{(1,0)}_b+g_{ab}\,V(\phi)\bigr)
        \Big] \nonumber\\
        &+\tfrac{F''(\phi)}{\alpha\,\mpl}\Big[2\,(\phi^{(1,0)})^2 R_{ab} - 4\,\phi^{(1,0)c}\, R_{c(a}\,\phi^{(1,0)}_{b)} \nonumber \\
        &\hspace{1.5cm}+
        \tfrac{4}{3}\, R \Big(\phi^{(1,0)}_a\,\phi^{(1,0)}_b    -  \tfrac{1}{4}\, g_{ab} (\phi^{(1,0)})^2\Big) - 4\, C_{acbd}\,\phi^{(1,0)c}\,\phi^{(1,0)d}
        \Big] \nonumber\\
        &+\tfrac{F'(\phi)}{\alpha\,\mpl}\Big[2\,\phi^{(0,1)}\, R_{ab} - 4\, R_{c(a} \nabla^{c}\phi^{(1,0)}_{b)} +
        \tfrac{4}{3} R\big( \nabla_{(a}\phi^{(1,0)}_{b)}  -  \tfrac{1}{4} g_{ab}\,\phi^{(0,1)}\big) \nonumber\\
        &\hspace{1.5cm}- 4\, C_{acbd} \nabla^{c}\phi^{(1,0)d} + V'(\phi)\left(2\,R_{ab}-\tfrac{1}{3}\,R\,g_{ab}\right)
        \Big] \nonumber\\
        &-\tfrac{4\,G''(\phi)}{\alpha\,\mpl}\,\tilde{C}_{acbd}\,\phi^{(1,0)c}\,\phi^{(1,0)d} 
        -\tfrac{4\,G'(\phi)}{\alpha\,\mpl}\Big[\epsilon_{cde(a}\,\phi^{(1,0)c}\,\nabla^dR_{b)}^{\phantom{b)}e}+\tilde C_{acbd}\,\nabla^c\phi^{(1,0)d} \Big] \nonumber\\
        &-\tfrac{H(\phi)}{2\,\alpha\,\mpl^4}\,(\phi^{(1,0)})^2\Big[\phi^{(1,0)}_a\,\phi^{(1,0)}_b-\tfrac{1}{4}\,g_{ab}\,(\phi^{(1,0)})^2\Big]\nonumber\\
        &+\tfrac{\gamma}{\alpha\,\mpl^2}\Big[
        2\,\phi^{(1,0)}_{(a}\,\nabla_{b)}\phi^{(0,1)} +g_{ab}\,\phi^{(0,1)}\left(\tfrac{1}{2}\,\phi^{(0,1)}+V'(\phi)\right)
        \Big]
        -\mathcal{G}_{ab}-\Box\mathcal{S}_{ab}\,,
        \label{eq:evo_Ric_bar}
\end{align}
\end{widetext}
with 
\begin{equation}
\begin{aligned}
    \Box\mathcal{S}_{ab}={}&\tfrac{1}{\mpl^2}\Big\{
    2\phi^{(1,0)}_{(a}\nabla_{b)}\phi^{(0,1)}
    +2\phi^{(1,0)}_{(a}R_{b)}{}^c\phi^{(1,0)}_c\\
    &+2\,V''(\phi)\phi^{(1,0)}_a\phi^{(1,0)}_b\\
    &+2(\nabla^c\phi^{(1,0)}_a)(\nabla_c\phi^{(1,0)}_b)
    -2\phi^{(1,0)}_{(a}\mathcal{D}_{b)}\\
    &+g_{ab}\Big[V''(\phi)\,\phi^{(1,0)}{}^{c}\phi^{(1,0)}_{c}\\
    &\hspace{1.2cm}+V'(\phi)\left(\phi^{(0,1)}+V'(\phi)\right)\Big] 
    \Big\}
\end{aligned}
\label{eq:eq_Sab}
\end{equation}
and
\begin{align}
    \mathcal{D}_a\equiv{}&\tfrac{1}{\ell}\,\gamma_{\rm C}\,n^b\nabla_b(\nabla_a\phi-\phi^{(1,0)}_a)\,,\\
    \mathcal{G}_{ab}\equiv{}&\tfrac{1}{\ell}\,\kappa_{\rm G}\,\bigl( 2\,n_{(a}\,\nabla^cG_{b)c}-\tfrac{1}{2}\,g_{ab}\,n^c\nabla^dG_{cd} \bigr)\,,
    \label{eq:residual_scalar_damping}
\end{align}
are the constraint damping terms that enforce $\nabla_a\phi-\phi^{(1,0)}_a=0$ and that the Einstein tensor $G_{ab}$ computed from dynamical variables $\bar R_{ab}$ and $\phi^{(1,0)}_a$ via \eqref{eq:ricci_residual_def} satisfies the Bianchi identity. Here $\gamma_C$ and $\kappa_G$ are dimensionless damping parameters, which serve to suppress constraint violating modes that might otherwise grow, damping them on time scales $1/\gamma_C$ and $1/\kappa_G$ respectively. In this paper, we typically set $\gamma_C \sim O(1)$ and $ \kappa_G \sim \ell/M$.

\begin{figure}[t!]
\includegraphics[width=\columnwidth]{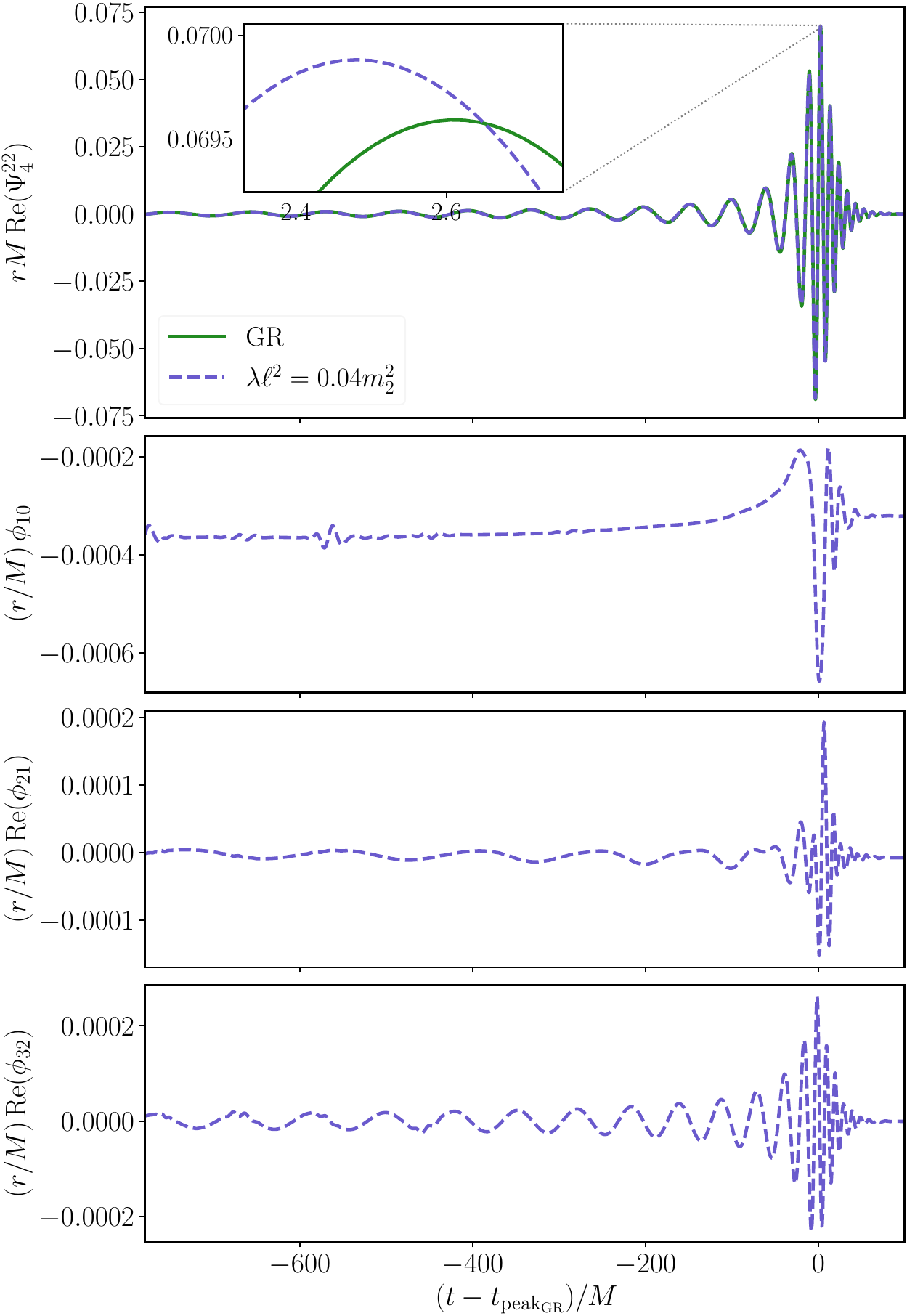} 
\caption{Gravitational and scalar waveforms from binary with mass ratio 1.22 and aligned dimensionless spins of magnitude 0.5, $\lambda\,\ell^2=0.04 m_2^2$ with $\ell^2=0.04M^2$, as well as GR for comparison: From top to bottom, we show $r\,M\,\text{Re}(\Psi_4^{22})$, $(r/M)\phi_{10}$, $(r/M)\text{Re}(\phi_{21})$ and $(r/M)\text{Re}(\phi_{32})$ extracted at $R=100M$. The time coordinate is shifted so that the GR waveform peaks at $t=0$. 
}\label{fig:inspiral_dCS_GHC}
\end{figure}

\PRLsection{{\bf Inspiral results for dCS gravity in GH coordinates}}
Here we present inspiral binaries in dCS gravity using the GHC implementation. 
We consider inspirals of BHs with mass ratio 1.22 and intrinsic dimensionless spins of 0.5 with both spins aligned with the orbital angular momentum. We choose an initial separation of $9M$ and fix the coupling parameter of the theory to be $\lambda\,\ell^2=0.04 m_2^2$, where $\ell^2=0.04 M^2$. Figure~\ref{fig:inspiral_dCS_GHC} shows the gravitational $(2,2)$ mode for dCS and GR, and the scalar $(1,0)$, $(2,1)$ and $(3,2)$ modes for dCS all extracted at $R=100M$. As explained in the main text, since the two BHs have aligned spins, there is a non-vanishing scalar dipole both before and after the merger.

\end{document}